\documentclass{iopart}

\usepackage{graphicx}
\usepackage{color}

\usepackage{iopams}

\def\rb{\boldsymbol{r}}
\def\rhor{\rho(\boldsymbol{r})}
\def\Vext{V_{\rm ext}(\boldsymbol{r})}
\def\taub{\tau_{\rm B}}
\def\aV{a_{\rm V}}
\def\aD{a_\Delta}

\begin{document}

\title{Compatibility waves drive crystal growth on patterned substrates}

\author{T~Neuhaus, M~Schmiedeberg, and H~L\"owen}
\address{Institut f{\"u}r Theoretische Physik II: Weiche Materie,
Heinrich-Heine-Universit\"at D\"usseldorf,
Universit\"atsstra{\ss}e 1, D-40225 D\"usseldorf, Germany}
\eads{tneuhaus@thphy.uni-duesseldorf.de}

\begin{abstract}
We explore the crystallization in a colloidal monolayer on a structured template starting from a few-particle nucleus. The competition between the substrate structure and that of the growing crystal induces a new crystal growth scenario. Unlike with the crystal growth in the bulk  where a well-defined and connected crystal-fluid interface grows into the fluid, we identify a mechanism where a "compatibility wave" of the prescribed nucleus with the underlying substrate structure dictates the growth direction and efficiency. The  growth process is strongly anisotropic and proceeds via transient island formation in front of an initial solid-fluid interface. We demonstrate the validity of this compatibility wave concept for a large class of substrate structures including a square-lattice and a quasicrystalline pattern. Dynamical density functional theory which provides a microscopic approach to the crystallization process is employed for colloidal hard spheres. Our predictions can be verified in experiments on confined colloids and  also bear consequences for molecular crystal growth on structured substrates.
\end{abstract}

\pacs{82.70.Dd, 64.70.D-, 05.20.Jj, 81.10.Aj}

\submitto{NJP}
\maketitle


\section{Introduction}
The novel material properties of ultrathin films promise fabrication of technologically relevant optical switching devices, high-density information storage media, and nanofilters with controlled porosity. In many cases, these films are crystalline monolayers which are grown on a patterned substrate acting as a template for solidification. Pivotal examples for these two-dimensional arrays range from sheets of graphene~\cite{Meyer2007,Kim2009} and organic molecules~\cite{Aizenberg1999,Briseno2006} to soft matter films composed of nanoparticles~\cite{Wang2004,Bigioni2006}, proteins~\cite{Meel2010}, polymers~\cite{Segalman2005} or colloidal particles~\cite{Blaaderen1997,Aizenberg2000,Reichhardt2002,Achim2008,Granato2011}.

There are various techniques to prepare crystalline layers on a structured substrate. Using heteroepitaxy from the gas phase~\cite{Shchukin1999}, crystalline islands are formed first on the substrate
which then expand until they merge to a covering layer while exhibiting at the same time layer-by-layer growth into the direction perpendicular to the substrate. Crystals on a patterned substrate can also be grown out of the liquid phase where typically a layer-by-layer growth is obtained perpendicular to the wall~\cite{Blaaderen1997,Heni2000,Toth2012,Dorosz2012,Jungblut2013}. A complementary technique uses
self-assembly within the monolayer by e.g.\ drying out the films~\cite{Bigioni2006,Aizenberg2000} or using electrophoretic deposition in the first place \cite{Trau1996} which basically corresponds to a two-dimensional crystallization process.

In order to control and steer the formation of crystalline sheets on a template a detailed understanding of the crystallization process on the scale of the individual particles  is necessary. Colloidal suspensions are excellent model systems to study the crystallization process on the particle scale~\cite{Ivlev2012}. A structured substrate can be realized by superimposing optical laser fields~\cite{Burns1990} which constrain the colloidal particles to a modulated external potential confining them to a two-dimensional layer~\cite{Bechinger2001a,Mikhael2008,Mikhael2010}. They thus offer the unique opportunity to observe the two-dimensional crystallization process in real-space.

Here, we explore the crystallization in a colloidal monolayer on a structured template starting from a few-particle nucleus. There is a crucial competition between the substrate structure and that one of the growing crystal (typically hexagonal for spherical particles) which gives rise to a new crystal growth scenario. Unlike with the crystal growth in the bulk~\cite{Langer1980} or on unstructured substrates~\cite{Toth2012} where a well-defined and connected crystal-fluid interface grows into the fluid possibly via faceting, branching or dendrite formation~\cite{Tegze2011}, we identify a mechanism where a ``compatibility wave'' of the prescribed nucleus with the underlying substrate structure dictates the growth direction and efficiency. The compatibility wave describes  the commensurability of the substrate structure with the stable hexagonal bulk crystal as documented in the Moir\'e pattern of the periodic solid induced by the imposed nucleus and the underlying substrate. Correspondingly, the growth process is strongly anisotropic and proceeds via transient island formation in front of an initial solid-fluid interface breaking the assumption of a well-defined single-connected interface topology. The compatibility wave concept is valid for a large variety of substrate structures e.g. for a square-lattice and a quasicrystalline pattern. Providing a microscopic approach to the crystallization process, dynamical density functional theory is employed for colloidal hard spheres. Our predictions can be verified by performing experiments on confined colloids and they also bear consequences for molecular crystal growth on structured substrates. Finally, by quenching the transient emerging crystal during growth it will be possible to fabricate remarkable extended and hollow crystal structures with new possible technological applications.

\section{Methods, model, and first results}
\begin{figure}[tb]
\centering
\includegraphics[width=0.95\textwidth]{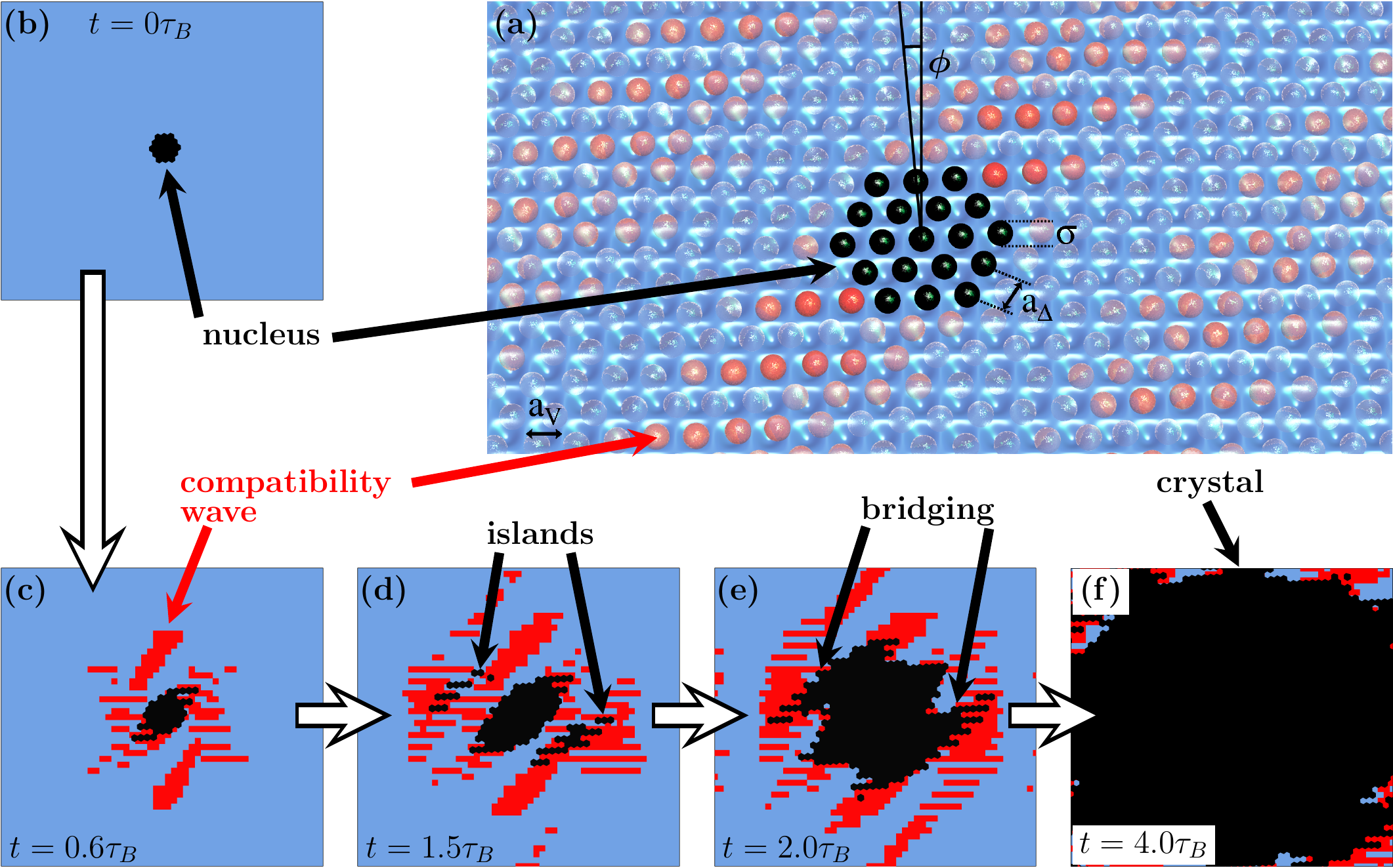}
\caption{\label{fig:scheme}(a) Schematic representation of a nucleus of hard disks with diameter $\sigma$ located on a two-dimensional substrate with square symmetry rotated counterclockwise by an angle $\phi=5^\circ$ relative to a symmetry direction of the substrate and lattice constant $\aV$. The snapshots (b)-(f) illustrate the growth of a spherical nucleus influenced by the substrate with amplitude $V_0=0.5k_BT$ at times (b) $t=0.0$, (c) $t=0.6\taub$, (d) $t=1.5\taub$, (e) $t=2.0\taub$, and (f) $t=4.0\taub$, where $\taub$ is the Brownian time. Red regions display the compatibility wave and are defined by density peaks above a threshold value $\rho_{\rm th}\sigma^2=1.5061$. Black regions denote crystalline areas ($\rho_{\rm th}\sigma^2=2.0$), whereas blue regions remain fluid ($\rho_{\rm th}\sigma^2\leq1.5060$). The growth process depicted in (b)-(f) is also shown in a movie that is part of the Supplemental Material~\cite{SI}.}
\end{figure}

We consider a system of hard disks of diameter $\sigma$ in the crystalline phase, influenced by an external potential with square symmetry. In figure~\ref{fig:scheme}(a), a schematic representation of the examined system is shown. A nucleus with inter-particle distances $\aD$ is placed on a square substrate with lattice constant $\aV$. The nucleus is rotated counter-clockwise by an angle $\phi$ with respect to a symmetry axis of the substrate. The fluid that surrounds the nucleus is modulated by the substrate potential. A typical time series of a growth process is shown in the snapshots (b)-(f) of figure~\ref{fig:scheme}. Movies of this and a similar growth process are available in the Supplemental Material~\cite{SI}. In contrast to the typical crystal growth where a well-defined crystal-fluid interface is growing, the external potential influences the dynamics such that regions which are compatible with the positions of the potential minima grow before regions that are less matching.

Influenced by the substrate potential, a modulation of the local density is induced. When the nucleus is positioned on the substrate, the hexagonal symmetry of the nucleus causes a modulation which starts growing but is different from the substrate symmetry. We call the superposition of these two modulations the ``compatibility wave'' which drives the growth process. The compatibility wave possesses maxima at the positions where the two patterns coincide.

In our schematic sketch in figure~\ref{fig:scheme}(a) the colour code indicates the distance of two nearest neighboring peaks of the crystal lattice sites and the substrate minima and therefore the compatibility positions are denoted by darker red particles. Snapshots (c)-(f) display the propagation of the compatibility wave in red whereas the grown crystal is displayed by black regions.

The external substrate potential $\Vext$ dictating the square symmetry is given by 
\begin{equation}
  \Vext=V_0\left[1-\frac{1}{4}\left(1-\cos(k_x x)\right)\left(1-\cos(k_y y)\right)\right],
 \label{eq:ext_pot}
\end{equation}
where $k_x=k_y=2\pi/\aV$ are the components of the reci\-procal lattice vector ${\bf k}$ and $V_0$ denotes the strength of the potential. We measured $V_0$ in units of the thermal energy $k_BT$, with temperature $T$ and Boltzmann's constant $k_B$, and chose an amplitude of $V_0=0.5k_BT$ and an area fraction of $\eta=0.74$. Furthermore, the length scale of the substrate $\aV$ is chosen such that on average there is always one particle per minimum. For these parameters, in equilibrium the system is in a triangular crystalline phase~\cite{Neuhaus2013}.

We use dynamical density functional theory (DDFT) for Brownian particles which is a dynamical generalization of classical density functional theory~\cite{Evans1979,Rosenfeld1989,Tarazona2008,Roth2010} and can be derived from the exact Smoluchowski equation~\cite{Marconi1999,Archer2004,Espanol2009}. The time-dependence of the density profiles $\rho(\rb,t)$ is given by a generalized diffusion equation
\begin{equation}
 \frac{\partial \rho(\rb,t)}{\partial t}=(k_BT)^{-1} D \nabla \cdot\left(\rho(\rb,t)\nabla\frac{\delta \Omega[T,A,\mu,\rho(\rb,t)]}{\delta \rho(\rb,t)}\right),
 \label{eq:DDFT}
\end{equation}
where $D$ is the short time diffusion coefficient. The grand canonical free energy $\Omega(T,A,\mu,[\rho(\rb,t)])$ is a functional of the time-dependent local density $\rho(\rb,t)$ and depends on temperature $T$, area of the system $A$, and the chemical potential $\mu$ which is used as a Lagrangian multiplier to fix the average particle number in the system. The grand canonical free energy can be split into the contribution of an ideal gas $\mathcal{F}_{\rm id}[\rhor]=k_BT\int {\rm d}\rb \rhor \left[\ln \left(\Lambda^2 \rhor \right)-1\right]$ including the (irrelevant) thermal wavelength $\Lambda$, the excess free energy $\mathcal{F}_{\rm exc}[\rhor]$ for which we use a recently developed approach from fundamental measure theory~\cite{Roth2012}, and $\mathcal{F}_{\rm ext}[\rhor]$ which describes the interaction of the particles with the substrate of \eref{eq:ext_pot} and which is given by
\begin{equation}
 \mathcal{F}_{\rm ext}[\rhor]=\int {\rm d}\rb\rhor[V_{\rm ext}(\rb)-\mu].
 \label{eq:Fext}
\end{equation}

In order to obtain the initial nucleus in figure~\ref{fig:scheme} and the movies we included in the Supplemental Material, a density profile modulated by the external potential is used. For a time of $0.07\taub$ an external pinning potential of Gaussian shape given by
\begin{equation}
 V_{\rm p}(\rb)=\sum \limits_i V_{\rm p}^{(0)}e^{-\alpha(\rb-\rb_i)^2}
\end{equation}
with a strong amplitude $V_{\rm p}^{(0)}/k_BT=4$ and width $\alpha\sigma^2=6$ is added to the external substrate potential so that the density peaks grow at the pinning positions $\rb_i$. After this time, the pinning potential is switched off and only the patterned substrate potential remains.

\section{Detailed analysis}
\begin{figure}[tb]
\centering
\includegraphics[width=0.65\textwidth]{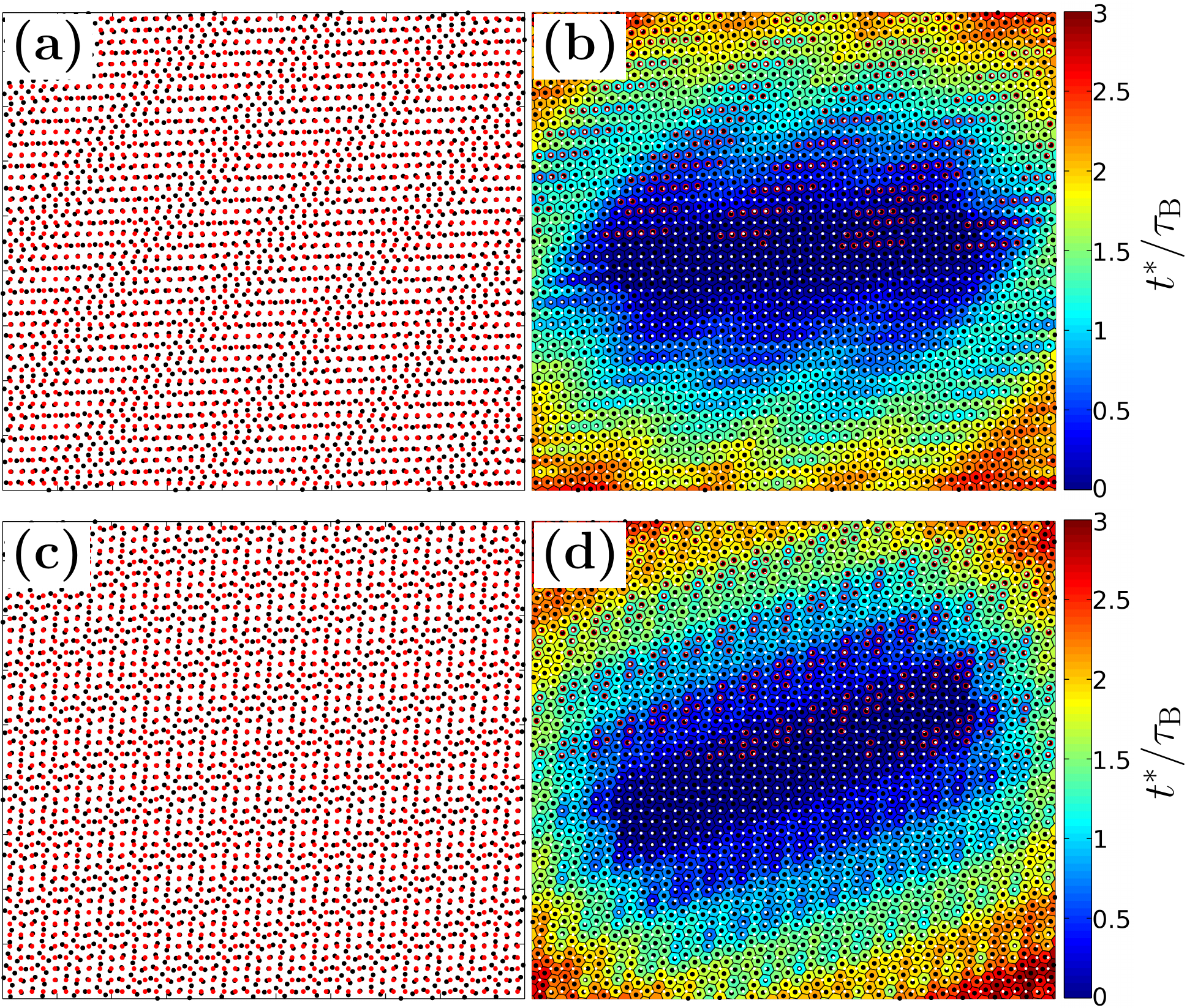}
\caption{\label{fig:stripe}A stripe of hard disks located on a square substrate at two different angles (a, b) $\phi=5^\circ$ and (c, d) $\phi=20^\circ$. The colour code in (b, d) displays the time $t^*/\taub$, at which the local density peak has reached a threshold value $\rho_{\rm th}\sigma^2=1.75$. The left plots (a) and (c) show the corresponding positions of the substrate minima (red) and the particle positions of the final crystal (black) as obtained from overlaying the crystal structure of the nucleus with the substrate structure (Moir\'e pattern). From these, compatibility regions can be extracted by measuring the distances between the points of both patterns.}
\vspace*{3pt}
\end{figure}

To reveal the underlying mechanism, we now focus onto a much simpler setup of planar growth and thus, we examine a nucleus consisting of a stripe of particles and analyze the growth perpendicular to the long side of the stripe. There are two different symmetries occurring -- the lattice sites of the crystal that is about to grow and the positions of the minima of the square potential. The superposition of these two patterns results in a Moir\'e structure which consists of regions where the crystal lattice sites almost coincide with the substrate wells and regions with less match. We call the regions of coincidence the compatibility regions which depend in width and orientation on the angle $\phi$. In figures~\ref{fig:stripe}(a, c) the patterns of crystal sites and the minima positions of the substrate potential are shown for two different values of $\phi$. At first glance, compatibility regions are visible, where the overlay of the patterns is less dense. The time resolved growth of the nucleus is shown in figures~\ref{fig:stripe}(b, d) for the same rotation angles $\phi$. The colour code denotes the times $t^*$ at which the local density peak has passed a threshold value $\rho_{\rm th}$. We obtain a growth behaviour strongly linked to the overlapping patterns with different symmetries where compatibility regions are clearly preferred during the growth process and thus, indicated by darker coloured patches in figures~\ref{fig:stripe}(b, d). At smaller angles $\phi$, the obtained pattern consists of broader stripes while at larger angles, these regions merge each other resulting in a pattern with thinner stripes. Thus, the islands that occur during the growth process are more pronounced for smaller angles $\phi$.

When the hexagonal crystal grows on the square substrate, the resulting crystal is no longer a perfect hexagonal one but it is slightly distorted. Thus, it becomes necessary to introduce a different lattice constant $\aD'$ which describes the obtained crystal more properly.

\begin{figure}[tb]
\centering
\includegraphics[width=0.5\textwidth]{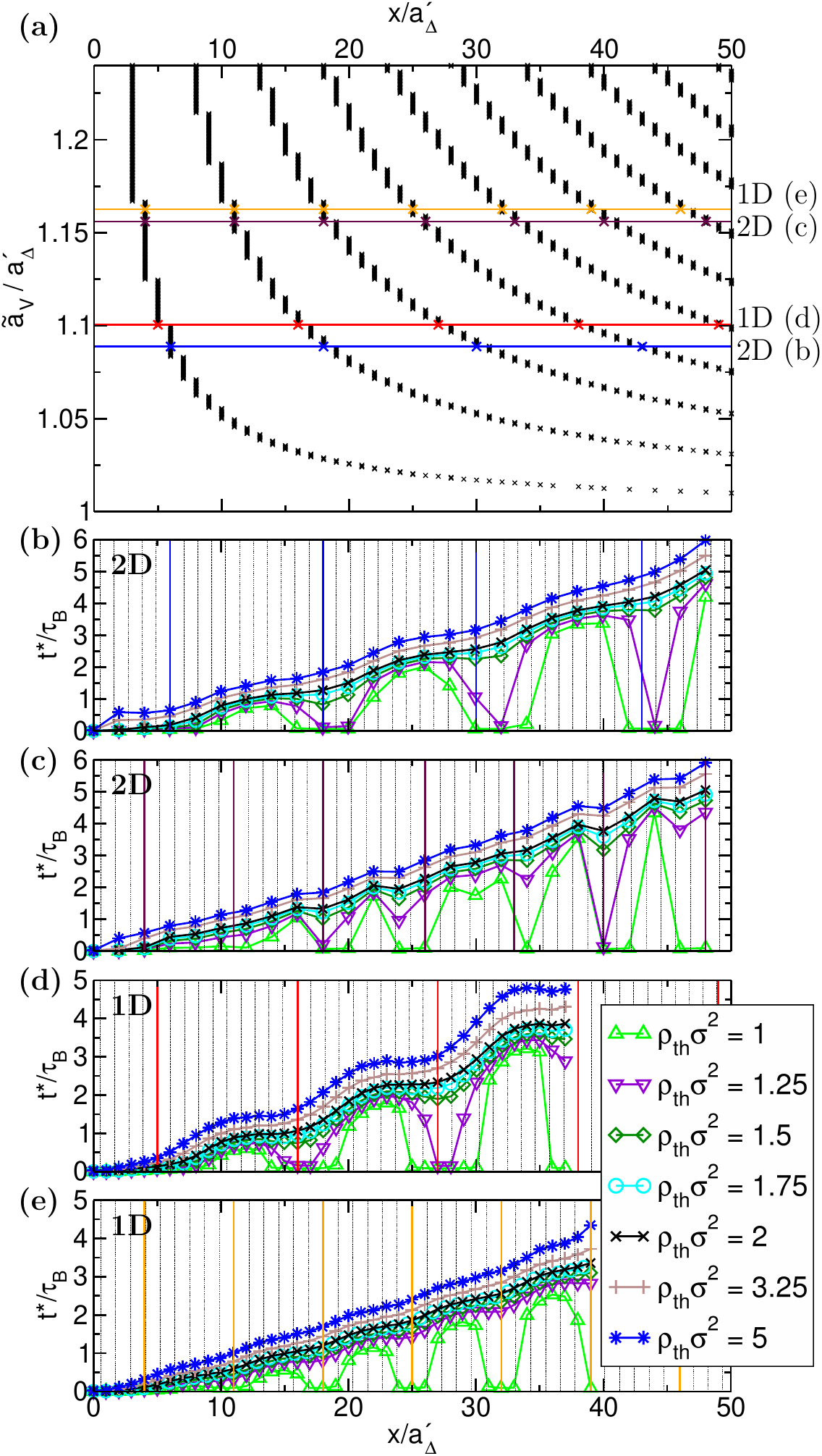}
\caption{\label{fig:time}Depending on the detune ratio of the lattice constants of the hexa\-gonal crystal $a'_\Delta$ and of the substrate potential $\tilde{a}_{\rm V}$ measured in the direction of growth, the discrete positions of best compatibility are plotted in (a). (b-e) Times $t^*$ at which the local density peaks of a growing crystal exceeds a given threshold value $\rho_{\rm th}$. The curves are obtained for a two-dimensional system (b, c) as well as for one-dimensional channels (d, e) as functions of the detune ratio $\tilde{a}_{\rm V}/\aD'$. Plots (b-e) are normalized to the lattice constant of the hexagonal crystal $a'_\Delta$, such that the best matching positions, indicated by coloured vertical lines, can be obtained from (a).}
\end{figure}

Figures~\ref{fig:time}(b, c) illustrate quantitative results of the growth in the direction perpendicular to the elongation of the nucleus. There are two length scales competing, namely the lattice constant of the external potential $\tilde{a}_{\rm V}$ measured in the direction of growth and the mean inter-particle distance of the nucleated crystal $\aD'$. Depending on the detune ratio of these length scales, $\tilde{a}_{\rm V}/\aD'$, compatibility regions can be found by comparing integer multiples of both length scales. Regarding the mismatch between these integer multiples of $\aD'$ and $\tilde{a}_{\rm V}$, local minima can be found which correspond to the predicted compatibility regions. In figure~\ref{fig:time}(a), the results of this analysis are shown. Here, the positions of the compatibility regions in $x$-direction are plotted for different detune ratios $\tilde{a}_{\rm V}/\aD'$. In the limit of $\tilde{a}_{\rm V}/\aD'=1$, all positions are compatible. However, for a very small detune the first compatibility region occurs at a very large distance $x$. With increasing the detune ratio, the distance to the first compatibility region as well as the distances between further compatibility regions decrease. Thus, compatibility regions occur more frequent, the larger the detune ratio is. In figure~\ref{fig:time}(b, c), the growth of the two-dimensional system is analyzed for two different rotation angles $\phi$ similar to those already shown in figure~\ref{fig:stripe}. The length scale of the external potential in the direction of growth, i.e. perpendicular to the stripe that is used as a nucleus, is given by
\begin{equation}
 \tilde{a}_{\rm V}(\phi)=\frac{\aV}{\cos(\phi)}.
\end{equation}

Accordingly, the positions of best compatibility for all parameters can be read from figure~\ref{fig:time}(a). From DDFT calculations, time-resolved density values $\rho(\rb,t)$ are obtained at all positions $\rb$. Based on these density distributions, a critical time $t^*$ can be extracted, at which the local density peak is above a certain threshold value $\rho_{\rm th}$. In figures~\ref{fig:time}(b, c), the values of the critical times $t^*$ are shown. For small rotation angles $\phi$, modulations in the growth behaviour are more pronounced, caused by a smaller mismatch of the two length scales. Thus, the compatibility positions, indicated by coloured vertical lines, appear at larger distances. For small $\rho_{\rm th}$, very short times are sufficient for the local density to pass the threshold value. The regions of short critical times $t^*$ match well to the proposed positions from figure~\ref{fig:time}(a). This can be understood as here the positions of the substrate potential coincide well with the crystal lattice sites. For higher threshold values, the curves denoting $t^*$ become monotonic. However, close to the compatibility positions the slope of the curves in the space-time plot is still much smaller than at less compatible positions.

This mechanism can also be examined in an effective one dimensional channel, where the external potential is simplified to a cosine-wave parallel to the growth direction, with lattice constant $\aV$ and amplitude $V_0$. For similar packing fractions $\eta$ and detune ratios $\aV/\aD'$, the time-resolved growth is explored and shown in figures~\ref{fig:time}(d, e). As the external potential affects the dynamics only in one direction compared with the two-dimensional growth scenario studied before, the shape of the plots becomes smoother. Still, regions of good compatibility coincide with the local minima of the critical times $t^*$ revealing the propagation of the compatibility wave. Regions of high compatibility grow earlier than those of lower compatibility.

\section{Substrates with other symmetries}
\begin{figure}[t]
\centering
\includegraphics[width=0.5\textwidth]{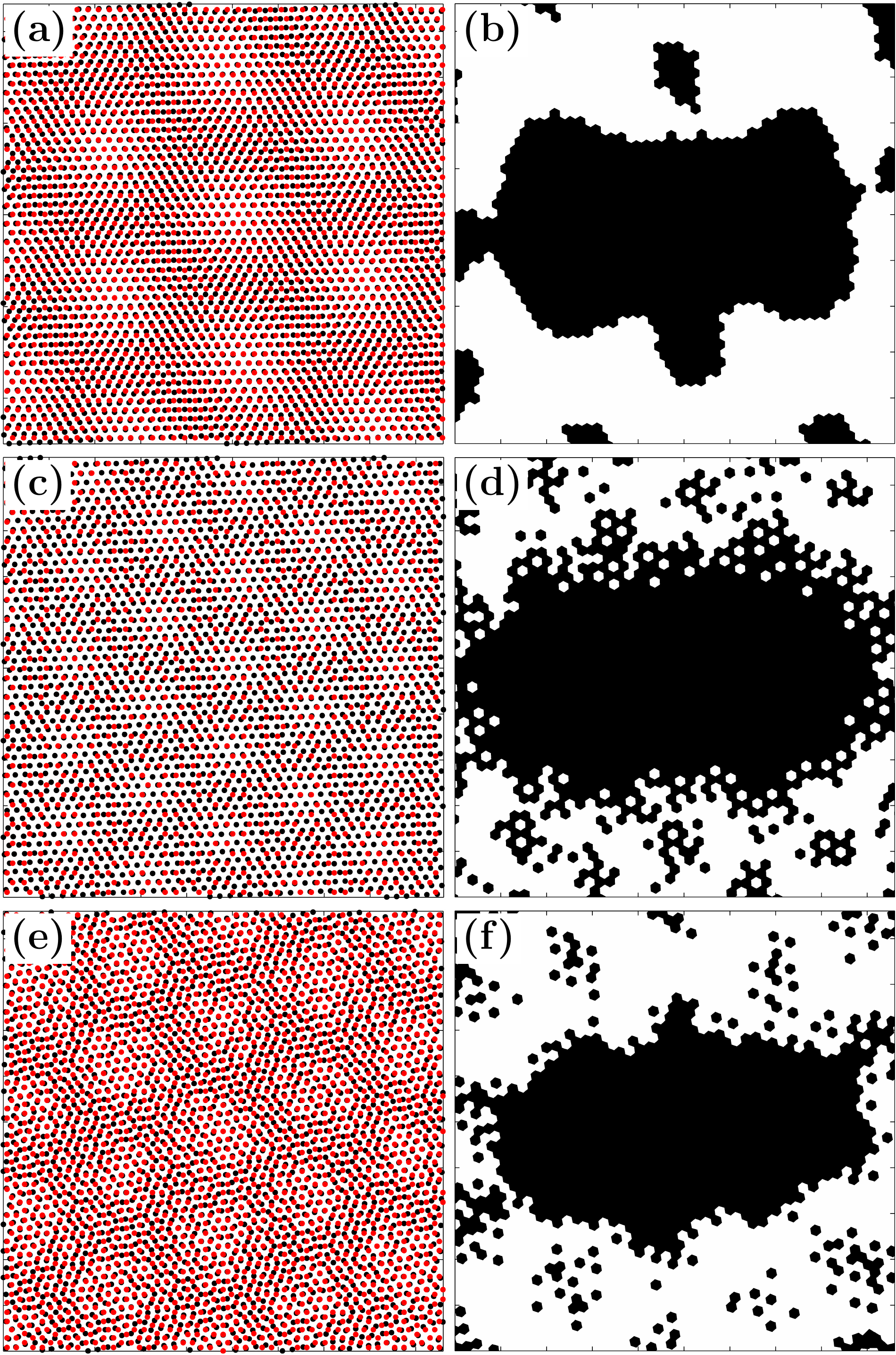}
\caption{\label{fig:pot}Superposition of a hexagonal lattice and the minima of a substrate potential (Moir\'e pattern) with (a) triangular, (c) honeycomb, (e) twelve-fold quasicrystalline symmetry. The detune ratio $\aV/\aD'$, as well as the rotation angle of the hexagonal lattice, are varied leading to three different combinations: (a) $\aV/\aD'=1.05$, $\phi=2^\circ$, (c) $\aV/\aD'=1.1$, $\phi=3^\circ$, and (e) $\aV/\aD'=1.0$, $\phi=10^\circ$. The quasicrystalline pattern is given by a square-triangle tiling (see, e.g.,~\cite{Hermisson1997}). (b, d, f) Snapshots of the growth of an initial stripe on the corresponding substrate potential with threshold values (b) $\rho_{\rm th}\sigma^2=1.1$, (d) $\rho_{\rm th}\sigma^2=1.0$, and (f) $\rho_{\rm th}\sigma^2=1.35$ at time $t=0.5\taub$. The system parameters in (b, f) are area fraction $\eta=0.73$, and interaction strength $V_0=0.1k_BT$ while in (d) the interaction strength is $V_0=0.2k_BT$.}
\vspace*{3pt}
\end{figure}
We expect that the growth scenario with compatibility waves is also important for a lot of other substrates with length scales that are incommensurate to the length scale of the growing crystal. In principle, the knowledge of the detune ratio is enough to predict the growth behaviour of a hexagonal crystal on any substrate. In figure~\ref{fig:pot}, three different types of two-dimensional substrates are overlaid with a rotated and detuned hexagonal lattice. In figure~\ref{fig:pot}(a), a hexagonal lattice is illustrated, leading to a superstructure with a six-fold symmetry of regions with better compatibility. The same symmetry occurs for a honeycomb lattice, displayed in figure~\ref{fig:pot}(c). Figure~\ref{fig:pot}(e) is created by employing a quasicrystalline lattice consisting of squares and triangles. Since a quasicrystal possesses at least two length scales per direction, the superstructure of compatibility regions is no longer periodic. For example, the square-triangle tiling that is used here (see, e.g.,~\cite{Hermisson1997}) has two incommensurate length scales per direction. The compatibility regions are given by the length scale of the growing lattice and one length scale of the quasicrystalline substrate, while the second length scales causes a modulation of the compatibility regions such that the compatibility regions never repeat in exactly the same way. Thus, the resulting pattern of a square-triangular and a hexagonal lattice shown in figure~\ref{fig:pot}(e) yields a six-fold symmetry but the repeating regions all differ slightly from each other. The growth of a nucleus with stripe symmetry on all three different substrate symmetries can be regarded and snapshots of this are shown in the corresponding figures~\ref{fig:pot}(b, d, f). Regions with density peaks above the threshold density are covered in black while white regions have density values below the threshold density. Obviously, the black spots apart from the crystal correspond to compatible regions of the two lattice structures.

As the growth process is similar in all presented cases, it can be extracted that for other kinds of substrate potentials, a crystal will grow in analogy to our prediction by a compatibility wave which drives the growth preferable at compatibility positions.

\section{Conclusions and Outlook}
In conclusion, we have investigated the dynamics of crystal growth for the case where a hexagonal nucleus of particles is placed on a patterned substrate. We found that due to the different length scales that occur in the direction of growth - one length scale given by the hexagonal crystal that is about to grow, and at least one other length scale by the substrate potential - no well-defined and connected crystal-fluid interface is obtained. In contrast, a compatibility wave drives the growth by favoring the regions of high compatibility of the two length scales. In principle, the same growth scenario also applies to the growth behaviour on more complex substrate potentials. The only relevant parameter that determines the position of the compatibility regions is the detune ratio of the two length scales. The compatibility wave concept is also relevant for molecular absorbate on atomic substrates. However, it is difficult to obersve directly for molecular systems.

A possible application of the compatibility wave concept is that a crystal is grown and impurities are confined at positions predefined by the compatibility regions. As islands form before the crystal-fluid interface, impurities can be locally trapped instead of being pushed in front of the interface. As a second application, it can be possible to infer the structure and the properties of the initial nucleus from the occurrence of the compatibility regions and the positions of the islands.

Furthermore, the growth behaviour of a nucleus on a patterned substrate in a three-dimensional system may be strongly related to the presented 2D results. Therefore, we believe that compatibility waves are important for a lot of systems where crystal growth processes on substrates with detuned length scales are studied and that the growth via a compatibility wave is a more suitable description than any scenario that requires a connected interface between the fluid and the crystalline phase.

\ack
We thank J. Horbach and M. Oettel for helpful discussions.
This work was supported by the DFG via SPP 1296 and SFB TR6.


\Bibliography{10}
\bibitem{Meyer2007} Meyer JC, Geim AK, Katsnelson MI, Novoselov KS, Booth TJ and Roth S 2007 The structure of suspended graphene sheets {\it Nature} {\bf 446} 60.
\bibitem{Kim2009} Kim KS, Zhao Y, Jang H, Lee SY, Kim JM, Kim KS, Ahn J-H, Kim P, Choi J-Y and Hong BH 2009 Large-scale pattern growth of graphene films for stretchable transparent electrodes {\it Nature} {\bf 457} 706.
\bibitem{Aizenberg1999} Aizenberg J, Black AJ and Whitesides GM, 1999 Control of crystal nucleation by patterned self-assembled monolayers {\it Nature} {\bf 398} 495.
\bibitem{Briseno2006} Briseno AL, Mannsfeld SCB, Ling MM, Liu S, Tseng RJ, Reese C, Roberts ME, Yang Y, Wudl F and Bao Z 2006 Patterning organic single-crystal transistor arrays {\it Nature} {\bf 444} 913.
\bibitem{Wang2004} Wang X, Summers CJ and Wang ZL 2004 Large-scale hexagonal-patterned growth of aligned zno nanorods for nano-optoelectronics and nanosensor arrays {\it Nano Lett.} {\bf 4} 423.
\bibitem{Bigioni2006} Bigioni TP, Lin X-M, Nguyen TT, Corwin EI, Witten TA and Jaeger HM 2006 Kinetically driven self assembly of highly ordered nanoparticle monolayers {\it Nat. Mater.} {\bf 5} 265.
\bibitem{Meel2010} van~Meel, JA, Sear, RP and Frenkel, D 2010 Design Principles for Broad-Spectrum Protein-Crystal Nucleants with Nanoscale Pits \PRL {\bf 105} 205501.
\bibitem{Segalman2005} Segalman RA 2005 Patterning with block copolymer thin films {\it Mat. Sci. Eng. R} {\bf 48} 191.
\bibitem{Blaaderen1997} Blaaderen A, Ruel R and Wiltzius P 1997 Template-directed colloidal crystallization {\it Nature} {\bf 385} 321.
\bibitem{Aizenberg2000} Aizenberg J, Braun PV and Wiltzius P 2000 Patterned colloidal deposition controlled by electrostatic and capillary forces \PRL {\bf 84} 2997.
\bibitem{Reichhardt2002} Reichhardt C and Olson CJ 2002 Novel Colloidal Crystalline States on Two-Dimensional Periodic Substrates, \PRL {\bf 88} 248301.
\bibitem{Achim2008} Achim CV, Karttunen M, Elder KR, Granato E, Ala-Nissila T and Ying SC 2008 Phase diagram of pinned lattices in the phase field crystal model {\it J. Phys.: Conf. Ser.} {\bf 100} 072001.
\bibitem{Granato2011} Granato E, Ramos JAP, Achim CV, Lehikoinen J, Ying SC, Ala-Nissila T and Elder KR 2011 Glassy phases and driven response of the phase-field-crystal model with random pinning, {\it Phys. Rev. E} {\bf 84} 031102.
\bibitem{Shchukin1999} Shchukin VA and Bimberg D 1999 Spontaneous ordering of nanostructures on crystal surfaces \RMP {\bf 71} 1125.
\bibitem{Heni2000} Heni M and L\"owen H 2000 Surface freezing on patterned substrates \PRL {\bf 85} 3668.
\bibitem{Toth2012} T\'oth GI, Tegze G, Pusztai T and Gr\'an\'asy L 2012 Heterogeneous crystal nucleation: The effect of lattice mismatch \PRL {\bf 108} 025502.
\bibitem{Dorosz2012} Dorosz S and Schilling T 2012 On the influence of a patterned substrate on crystallization in suspensions of hard spheres \JCP {\bf 136} 044702.
\bibitem{Jungblut2013} Jungblut S and Dellago C 2013 Crystallization on prestructured seeds {\it Phys. Rev. E} {\bf 87} 012305.
\bibitem{Trau1996} Trau M, Saville DA and Aksay IA 1996 Field-induced layering of colloidal crystals {\it Science} {\bf 272} 706.
\bibitem{Ivlev2012} Ivlev A, Morfill G and L\"owen H 2012 {\it Complex Plasmas and Colloidal Dispersions: Particle-Resolved Studies of Classical Liquids and Solids, Series in Soft Condensed Matter Vol. 5} ed Andelmann D, Reiter G, (World Scientific).
\bibitem{Burns1990} Burns MM, Fournier J-M and Golovchenko JA 1990 Optical matter: Crystallization and binding in intense optical fields {\it Science} {\bf 249} 749.
\bibitem{Bechinger2001a} Bechinger C, Brunner M and Leiderer P 2001 Phase behavior of two-dimensional colloidal systems in the presence of periodic light fields \PRL {\bf 86} 930.
\bibitem{Mikhael2008} Mikhael J, Roth J, Helden L and Bechinger C 2008 Archimedean-like tiling on decagonal quasicrystalline surfaces {\it Nature} {\bf 454} 501.
\bibitem{Mikhael2010} Mikhael J, Schmiedeberg M, Rausch S, Roth J, Stark H and Bechinger C 2010 Proliferation of anomalous symmetries in colloidal monolayers subjected to quasiperiodic light fields, {\it Proc. Natl. Acad. Sci. USA} {\bf 107} 7214.
\bibitem{Langer1980} Langer JS 1980 Instabilities and pattern formation in crystal growth \RMP {\bf 52} 1.
\bibitem{Tegze2011} Tegze G, T\'oth GI and Gr\'an\'asy L 2011 Faceting and branching in 2d crystal growth \PRL {\bf 106} 195502.
\bibitem{SI}{See the Supplemental Material at URL for movies of the crystal growth.}
\bibitem{Neuhaus2013} Neuhaus T, Marechal M, Schmiedeberg M and L\"owen H 2013 Rhombic Preordering on a Square Substrate \PRL {\bf 110} 118301.
\bibitem{Hermisson1997} Hermisson J, Richard C and Baake M 1997 A guide to the symmetry structure of quasiperiodic tiling classes {\it J. Phys. I France} {\bf 7} 1003.
\bibitem{Evans1979} Evans R 1979 The nature of the liquid-vapour interface and other topics in the statistical mechanics of non-uniform, classical fluids {\it Adv. Phys.} {\bf 28} 143.
\bibitem{Rosenfeld1989} Rosenfeld Y 1989 Free-energy model for the inhomogeneous hard-sphere fluid mixture and density-functional theory of freezing \PRL {\bf 63} 980.
\bibitem{Tarazona2008} Tarazona P, Cuesta JA and Martinez-Raton Y 2008 Density functional theories of hard particle systems {\it Lect. Notes Phys.} {\bf 753} 247.
\bibitem{Roth2010} Roth R 2010 Fundamental measure theory for hard-sphere mixtures: a review \JPCM {\bf 22} 063102.
\bibitem{Marconi1999} Marconi UMB and Tarazona P 1999 Dynamic density functional theory of fluids \JCP {\bf 110} 8032.
\bibitem{Archer2004} Archer AJ and Evans R 2004 Dynamical density functional theory and its application to spinodal decomposition \JCP {\bf 121} 4246.
\bibitem{Espanol2009} Espanol P and L\"owen H 2009 Derivation of dynamical density functional theory using the projection operator technique \JCP {\bf 131} 244101.
\bibitem{Roth2012} Roth R, Mecke K and Oettel M 2012 Communication: Fundamental measure theory for hard disks: Fluid and solid \JCP {\bf 136} 081101.
\endbib
\end{document}